\documentclass[12pt,a4paper]{conference}

\usepackage{fancyhdr}
\usepackage{graphicx,amsmath,amssymb,cite}
\usepackage{multind}
\makeindex{author} \makeindex{subject}

\newcommand{\beq}{\begin{equation}}
\newcommand{\eeq}[1]{\label{#1}\end{equation}}
\newcommand{\eeqn}{\end{equation}}

\newcommand{\beqa}{\begin{eqnarray}}
\newcommand{\eeqa}[1]{\label{#1}\end{eqnarray}}
\newcommand{\eeqan}{\end{eqnarray}}

\let\bar=\overbar

\newcommand{\Dslash}{\not{\hbox{\kern-4pt $D$}}}
\newcommand{\dslash}{\not{\hbox{\kern-2pt $\del$}}}

\newcommand{\msb}{{\bar{\ssstyle M \kern -1pt S}}}

\begin{document}

\Chapter{Lattice QCD study of $g_A^{N^*N^*}$ 
with two flavors of dynamical quarks}
{}{T. T. Takahashi and T. Kunihiro}

\addcontentsline{toc}{chapter}{{\it T. T. Takahashi and T. Kunihiro}} \label{authorStart}

\begin{raggedright}

{\it T. T. Takahashi and T. Kunihiro }\index{author}{T. T. Takahashi and
 T. Kunihiro}\\
Yukawa Institute for Theoretical Physics\\
Kyoto University\\
Kitashirakawa-Oiwakecho, Sakyo, Kyoto 606-8502, Japan
\bigskip\bigskip

\end{raggedright}

\begin{center}
\textbf{Abstract}
\end{center}

We report the first lattice QCD result of 
the axial charge of N(1535), $g_A^{N^*N^*}$.
The measurement is performed with two flavors of dynamical quarks
employing the renormalization-group improved gauge action at $\beta$=1.95
and the mean-field improved clover quark action 
with the hopping parameters, $\kappa$=0.1375, 0.1390 and 0.1400.
In order to avoid the signal contaminations 
by N(1650) lying just 100 MeV above N(1535),
we construct 2$\times$2 correlation matrices and diagonalize them
so that clear signal separation can be found.
The wraparound contributions in the correlator,
which can be another source of signal contamination,
are eliminated by imposing the Dirichlet boundary condition
in the temporal direction.
We find that the axial charge of N(1535) takes small values as
$g_A^{N^*N^*}\sim 0.2$, independent of quark masses,
in the pion-mass range of 0.7 to 1.1 GeV.

\section{Introduction}

Chiral symmetry is an approximate global symmetry in 
Quantum ChromoDynamics (QCD),
and the symmetry and its spontaneous breaking
is one of the key ingredients in the low-energy hadron physics.
For instance, all the hadrons can be classified
into representations of $SU(N_f)_L\times SU(N_f)_R$.
Once we fix the representations, 
it gives strong constraints to low-energy effective lagrangians 
and possible terms are uniquely determined besides overall constants.
To embody chiral symmetry in effective lagrangians,
we have two famous ways; the linear and the non-linear representations.
The non-linear representation has been well studied and successful
especially in the context of the chiral perturbation theory.
The linear representation
with scalar mesons as chiral partners of Nambu-Goldstone bosons
would be important around the chiral restoration point
at high temperature/density.

As for the realization of chiral representations in the baryon sector
in the linear representation, there could be naively two 
ways~\cite{DeTar:1988kn,Jido:1998av}.
One is the naive assignment 
and the other is the so-called mirror assignment introduced 
by DeTar and Kunihiro~\cite{DeTar:1988kn}.
We can find several important differences
between these two assignments
in the couplings or in the nucleon masses.
For example,
the nucleon and its parity partner belong to the same chiral multiplet
and there can exist chirally-invariant mass terms of nucleons
in the mirror assignment~\cite{DeTar:1988kn}.
Due to the mass terms, nucleons can be massive 
even when the chiral condensate takes a small value or zero,
whereas nucleon masses are simply proportional to the chiral condensate
in the naive assignment~\cite{DeTar:1988kn,Jido:1998av},
which would be the most important difference
between the naive and mirror cases.
Such differences play crucial roles
at finite temperature/density systems
and it should be revealed directly from QCD.
In order to clarify which assignment is natural,
it would be advantageous to measure the axial charge of N(1535),
which we assume as the chiral partner of N(940),
because the axial charges of N(940) and N(1535) 
are sensitive to the chiral structure of baryons~\cite{DeTar:1988kn,Jido:1998av}
and have the same (different) signs in the naive (mirror) assignments.

In this report, we show the first unquenched lattice QCD study
of $g_A^{N^*N^*}$ as well as $g_A^{NN}$.
(For the details, see~\cite{TakahashiKunihiro}.)
We employ $16^3\times 32$ lattice with two flavors of dynamical quarks,
generated~\cite{AliKhan:2001tx} with the renormalization-group improved
gauge action at $\beta=1.95$ and the mean field improved clover quark action
with the clover coefficient $c_{\rm SW}=1.530$.
The calculations are done with the hopping parameters,
$\kappa_{\rm sea},\kappa_{\rm val}=0.1375$, 0.1390 and 0.1400.

\section{Lattice QCD formulations and results}

N(1535)is the ground-state nucleon in $\frac12^-$channel.
Though a ground state signal 
can be in principle isolated using a large Euclidean time separation
between the source and the sink points in correlators,
we could suffer from the signal contamination by N(1650)
lying just 100 MeV above.
With the aim to separate the signals in a proper way
and to optimize operators, we diagonalize correlation matrices
constructed with two independent operators;
$
N_1(x)\equiv \varepsilon_{\rm abc}u^a(x)(u^b(x)C\gamma_5 d^c(x)),
$
$
N_2(x)\equiv \varepsilon_{\rm abc}\gamma_5 u^a(x)(u^b(x)C d^c(x)).\nonumber
$
Here, $u(x)$ and $d(x)$ are the Dirac spinors for u- and d- quarks,
respectively, and a,b,c denote the color indices.
We eliminate wraparound effects,
which could be another possible sources of contamination,
imposing the Dirichlet boundary condition in the temporal direction.

With the optimized operators ${\mathcal N(x)}$, 
we can obtain vector(axial) charges 
$g_V$($g_A$) as follows.
\begin{equation}
g_V \rightarrow
\frac{
{\rm tr}
\gamma_4
\Gamma
\langle {\cal N}(t_{\rm snk})
V_4(t)
\overline{{\cal N}}(t_{\rm src})\rangle
}{
{\rm tr}
\Gamma \langle {\cal N}(t_{\rm snk})
\overline{{\cal N}}(t_{\rm src})\rangle
}
\quad\quad
(t_{\rm snk}\gg t\gg t_{\rm src})
\label{3pf1}
\end{equation}
and
\begin{equation}
g_A \rightarrow
\frac{
{\rm tr}
\gamma_5\gamma_3
\Gamma
\langle {\cal N}(t_{\rm snk})
A_3(t)
\overline{{\cal N}}(t_{\rm src})\rangle
}{
{\rm tr}\Gamma \langle {\cal N}(t_{\rm snk})
\overline{{\cal N}}(t_{\rm src})\rangle
}
\quad\quad
(t_{\rm snk}\gg t\gg t_{\rm src}),
\label{3pf2}
\end{equation}
with $\Gamma\equiv \frac{1+\gamma_4}{2}$.
Here, 
$A_\mu(t)\equiv \sum_{\bf x}
\bar u(x)\gamma_\mu\gamma_5 u(x)
-\bar d(x)\gamma_\mu\gamma_5 d(x)$
and
$V_\mu(t)\equiv \sum_{\bf x}
\bar u(x)\gamma_\mu u(x)
-\bar d(x)\gamma_\mu d(x)$
are the zero-momentum projected axial and vector currents,
and the traces are taken over spinor indices.
All the unwanted quantities,
such as the normalization factors,
are all canceled out between the denominator and the numerator.


\begin{figure}[h]
\begin{center}
\includegraphics[scale=0.37]{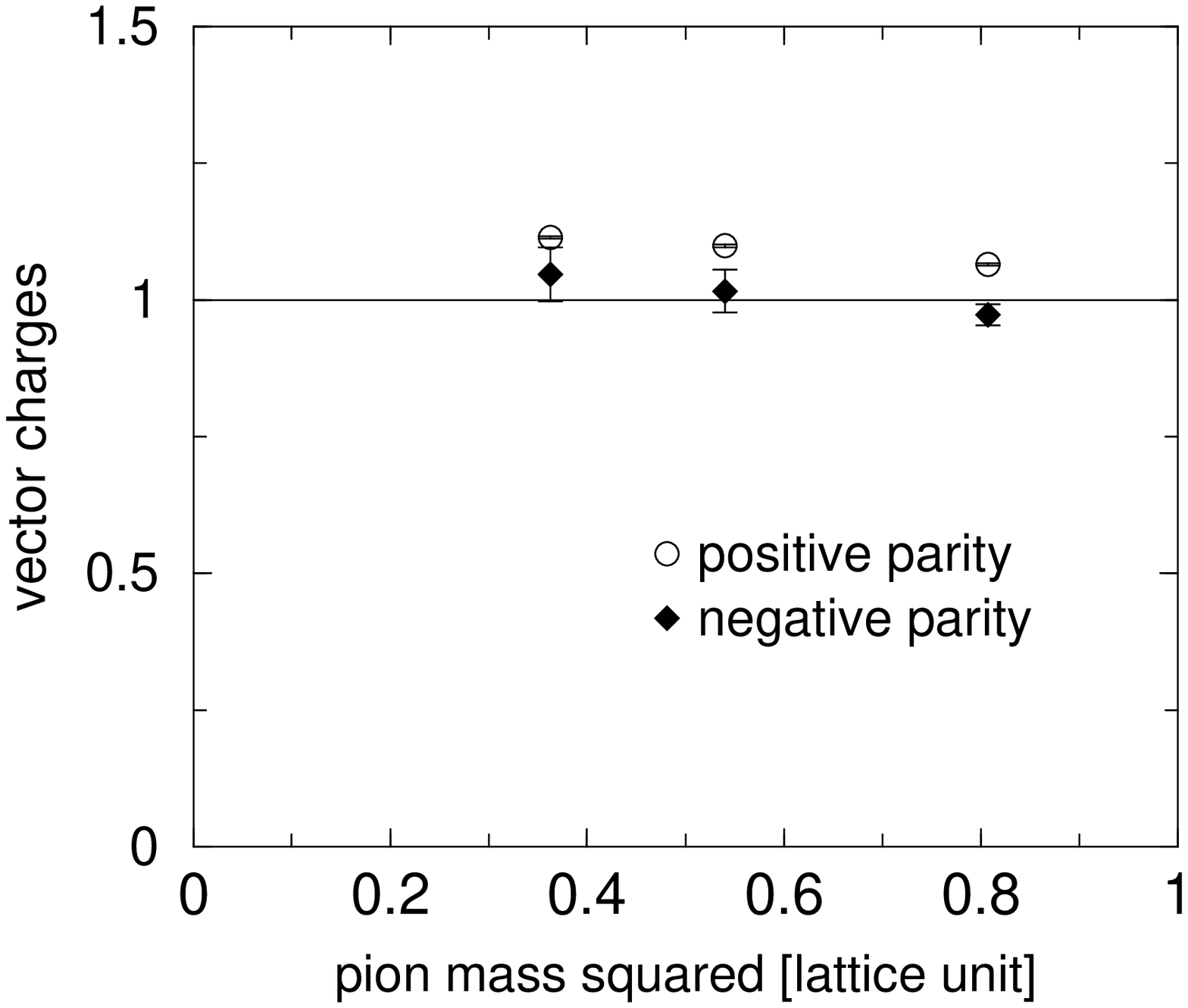}
\includegraphics[scale=0.37]{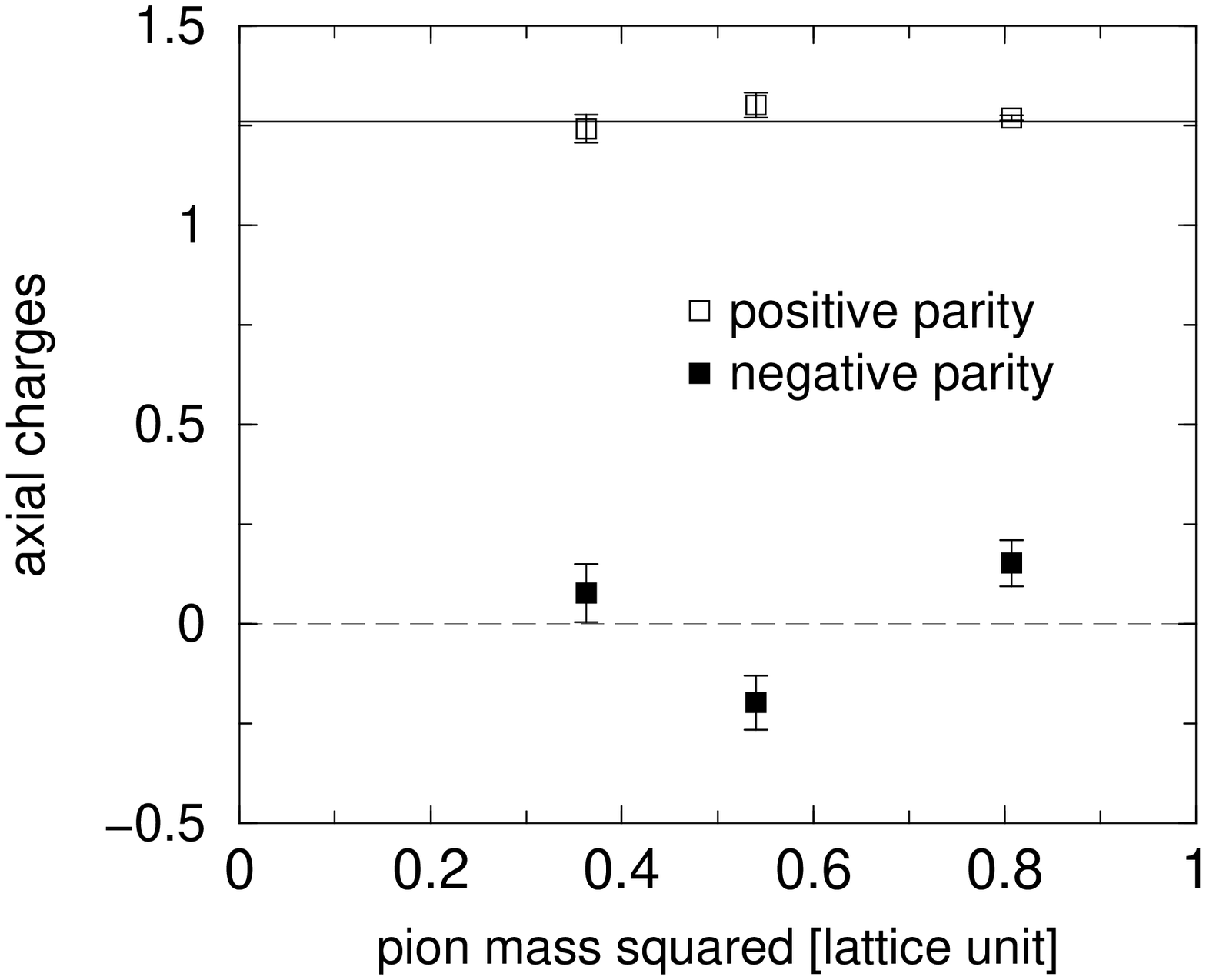}
\caption{
\label{finalcharges}
The renormalized vector and axial charges of the positive- and the 
negative-parity nucleons are plotted
as the function of the squared pion mass $m_\pi^2$.
The left panel shows the results of the vector charges
and the right panel the results of the axial charges.
In the left panel, the solid line is drawn at 
${g}_V=1$ for reference.
In the right panel, the solid line is drawn at 
${g}_A=1.26$ and the dashed line is drawn at 
${g}_A=0$.
}
\end{center}
\end{figure}

The renormalization factors for bilinear operators
are determined with the constants listed in Ref.~\cite{AliKhan:2001tx}.
We plot in the left panel in Fig.~\ref{finalcharges}
the vector charges of the positive- and the negative-parity nucleons,
which should be unity.
The open (filled) symbols denote the vector charges
of the positive- (negative-) parity nucleon at each hopping parameter.
We can find about 10\% deviations from unity,
which can be considered to come from
the systematic errors in the renormalization factors.
We should then take into account at least 10\% systematic errors
in our results.
The axial charges of the positive-parity nucleon
at each hopping parameter are plotted in the right panel.
They are shown as the open symbols.
One can find the good agreement between
the lattice data and 1.26, the experimental value.

We finally show the axial charges of the negative-parity nucleon
in the right panel.
One finds at a glance that they take quite small values,
as $g_A^{N^*N^*}\sim 0.2$
and that even the sign is quark-mass dependent.
While the wavy behavior might come from
the sensitiveness of $g_A^{N^*N^*}$ to quark masses,
this behavior may indicate that
$g_A^{N^*N^*}$ is rather consistent with zero.
The small $g_A^{N^*N^*}$ reflects
the interesting chiral structure of 
baryons~\cite{DeTar:1988kn,Jido:1998av,Jido:1999hd,Jaffe:2006jy,Glozman:2007ek}.

The present quark masses are unfortunately so heavy
that their related pion masses are 700MeV $\sim$ 1.1GeV.
In order to reveal the chiral structure,
much lighter u,d quarks are indispensable.
The study of the axial charge of Roper or N(1650)
as well as the inclusion of strange sea quarks could also 
cast light on the low-energy chiral structure of baryons.
They are left for further study.

\section*{Acknowledgments}

All the numerical calculations were performed
with NEC SX-8 at CMC, Osaka university and at YITP, Kyoto university.
The unquenched gauge configurations
employed in our analysis
were all generated by CP-PACS collaboration~\cite{AliKhan:2001tx}.




\end{document}